\documentclass[review,1p,times,twocolumn]{elsarticle}

\usepackage{graphicx}

\usepackage{amssymb}
\usepackage{amsmath}

\usepackage{wasysym}
\usepackage{import}
\usepackage{color}
\usepackage[colorlinks,urlcolor=blue,bookmarks=false]{hyperref}
\usepackage{booktabs}
\usepackage{symbol_short_cuts}
\usepackage{xfrac}
\usepackage[british]{babel}
\usepackage{array}
\usepackage{subcaption}
\usepackage[T1]{fontenc}
\usepackage[utf8]{inputenc}
\usepackage[normalem]{ulem} \newcommand\redsout{\bgroup\markoverwith{\textcolor{red}{\rule[0.5ex]{2pt}{0.4pt}}}\ULon} 

\usepackage[nodots]{numcompress}

\usepackage{lineno}

\newcommand{\overbar}[1]{\mkern 1.5mu\overline{\mkern-1.5mu#1\mkern-1.5mu}\mkern 1.5mu}
\newcommand{\obar}[2]{\mkern 1.5mu\overline{\mkern-1.5mu#2\mkern-#1mu}\mkern #1mu}

\usepackage[svgnames,rgb]{xcolor}
\usepackage[final,hoffset=-4mm,color=yellow,markup=Highlight]{pdfcomment} 
\makeatletter
\let\oldmarginnote\marginnote
\renewcommand*{\marginnote}[1]{%
	\begingroup%
	\ifodd\value{page}
	\if@firstcolumn\reversemarginpar\fi
	\else
	\if@firstcolumn\else\reversemarginpar\fi
	\fi
	\oldmarginnote{#1}%
	\endgroup%
}
\makeatother

\journal{Renewable Energy}

\begin{document}

\begin{frontmatter}



\title{Airborne Wind Energy Resource Analysis}
\marginnote{\pdfcomment{
[Remark] Suggestions for potential reviewers (priority decreasing):\textCR
1. Christina Archer <carcher@udel.edu>\textCR
2. Ilona Bastigkeit <ilona.bastigkeit@iwes.fraunhofer.de>\textCR
3. Gerrit Wolken-Möhlmann <gerrit.wolken-moehlmann@iwes.fraunhofer.de>\textCR
4. Ken Caldeira <kcaldeira@carnegie.stanford.edu>\textCR
5. Johan Meyers <johan.meyers@kuleuven.be>\textCR
6. Markus Sommerfeld <msommerf@uvic.ca>
}}


\author[a]{Philip Bechtle}\ead{bechtle@physik.uni-bonn.de}
\author[b]{Mark Schelbergen}\ead{m.schelbergen@tudelft.nl}
\author[b]{Roland Schmehl\corref{cor1}}\ead{r.schmehl@tudelft.nl}
\author[c]{Udo Zillmann}\ead{zillmann@airbornewindeurope.org}
\author[b]{Simon Watson}\ead{s.j.watson@tudelft.nl}
\address[a]{University of Bonn, Physikalisches Institut, Germany}
\address[b]{Delft University of Technology, Faculty  of Aerospace Engineering, Delft, Netherlands}
\address[c]{Airborne Wind Europe, Brussels, Belgium}
\cortext[cor1]{Corresponding author. Tel.: +31 15 278 5318}

\begin{abstract}
We compare the available wind resources for conventional wind turbines
and for airborne wind energy systems. Accessing higher altitudes and
continuously adjusting the harvesting operation to the wind resource
substantially increases the potential energy yield. The study is based 
on the ERA5 reanalysis data which covers a period of 7 years with hourly
estimates at a surface resolution of 31 $\times$ 31 km and a vertical resolution 
of 137 barometric altitude levels. We present detailed wind statistics for a location 
in the English Channel and then expand the analysis to a surface grid of 
Western and Central Europe with a resolution of 110 $\times$ 110 km. Over 
the land mass and coastal areas of Europe we find that compared to a fixed 
harvesting height at the approximate hub height of wind turbines, the 
wind power density which is available for 95\,\% of the time increases by a factor 
of two.
\end{abstract}


\end{frontmatter}



\section{Introduction}
\label{sec:introduction}

Previous resource assessments have shown that the wind energy available in the atmosphere could theoretically power the world \cite{Archer2005}. The precise extend of this energy potential is however still a subject of scientific debate. Uncertain is, for example, what effect a large-scale energy extraction would have on the overall resource and how the vertical energy exchange between the wind layers would influence an extraction on this scale. \citet{Miller2011} estimate a maximum of 18 to 68 TW that can be extracted over land. \citet{Jacobson2012} include also coastal ocean regions outside Antarctica and revise the saturation wind power potential to 80 TW. \citet{Adams2013} use a mesoscale model and predict that the energy potential is significantly lower, at 20 TW, a result that is roughly confirmed by \citet{Miller2015}. On the other hand, \citet{Emeis2018} estimates the total extractable wind power potential to be about 61 TW. \citet{Dupond2018} review these estimates and conclude that the global wind energy potential is substantially lower than previously established when both physical limits and a high cut-off value for the energy returned on energy invested (EROI > 10) is applied.

\begin{figure}[h!]
	\centering
	\includegraphics[width=\textwidth]{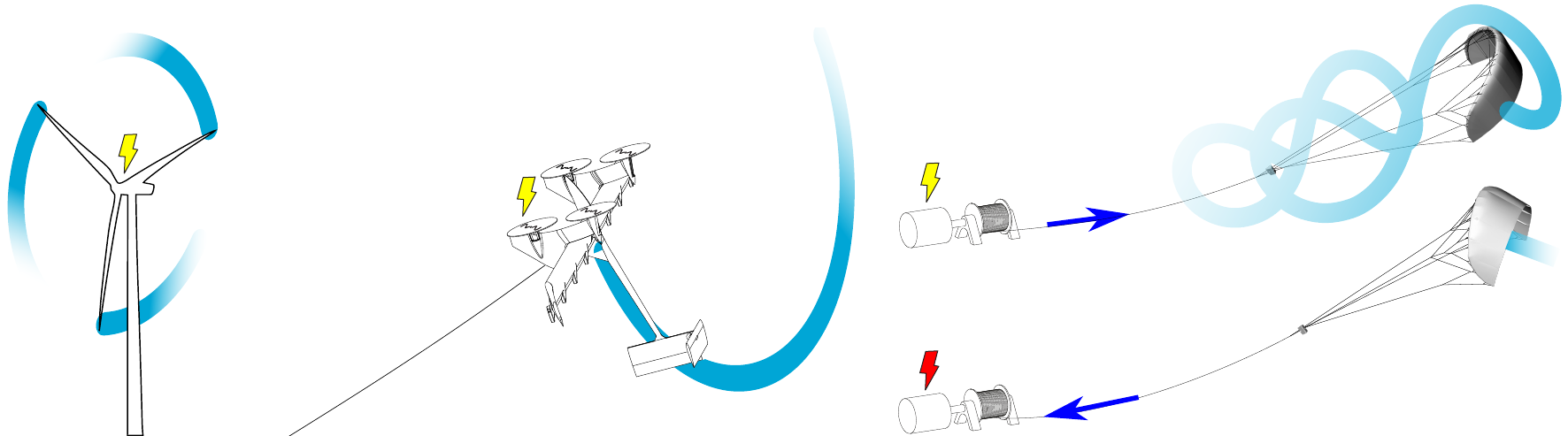}
	\caption{AWE systems replace the tips of a conventional wind turbine (left) by tethered flying devices. Electricity is generated either continuously onboard and transmitted to the ground by a conducting tether (center), or on the ground, with the tether transmitting the alternating mechanical power of a pumping cycle (right).}
	\label{fig:concepts}
\end{figure}

Most of these studies account only for an energy extraction close to the surface, using conventional wind turbines. Airborne wind energy (AWE) systems use tethered flying devices to access higher altitudes where wind is generally stronger and more persistent. The combination of cost savings due to lower material consumption and the ability to continuously adjust the harvesting height to the wind conditions bears the potential to substantially decrease the cost of energy and to access an energy resource that has not been used so far~\cite{Zillmann2018}. The two different AWE conversion concepts are illustrated in Figure~\ref{fig:concepts} next to a conventional horizontal axis wind turbine (HAWT).
A majority of the currently pursued development projects aims at ground-based conversion employing crosswind operation in a pumping cycle~\cite{Cherubini2015,Diehl2017,Schmehl2018}. Technological challenges are the reliability and robustness of the flying systems~\cite{Friedl2015,Damas2017}, reducing the land surface use~\cite{Faggiani2018} and the regulatory framework~\cite{Meister2016,Salma2018}.%

\begin{figure}[h!]
	\centering
	\includegraphics[width=\textwidth]{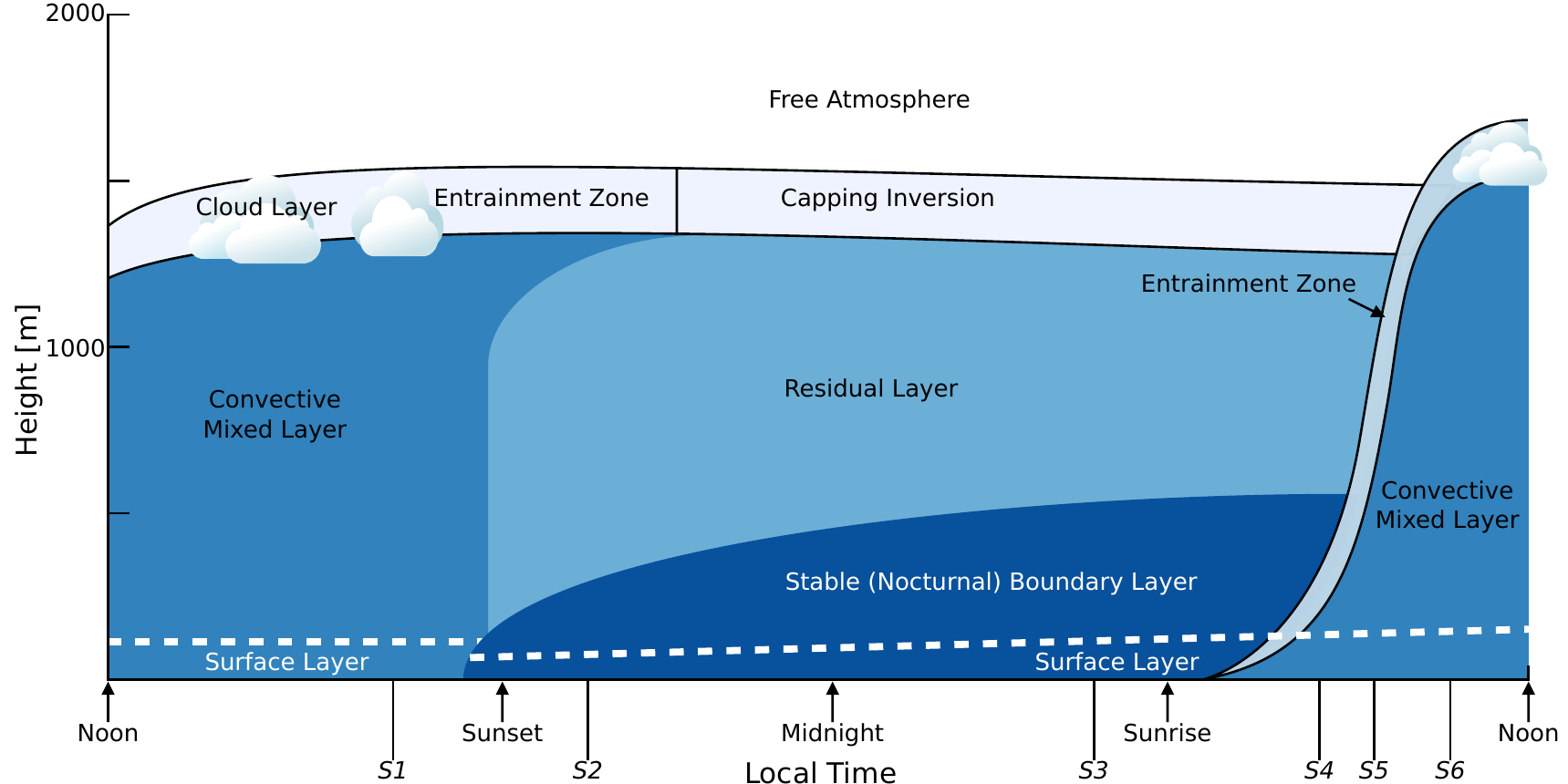}
	\caption{Temporal evolution of different regions of the atmospheric boundary layer in high pressure regions over land (adapted from \citet{Stull1988}).}
	\label{fig:atmospheric_boundary_layer}
\end{figure}

Because they harvest energy at heights beyond the reach of conventional tower-based wind turbines, AWE systems are exposed to different regions of the atmospheric boundary layer. The evolution of these regions along a day is illustrated schematically in Figure~\ref{fig:atmospheric_boundary_layer}.
The boundary layer flow is driven by the geostrophic wind at 1000 -- 1500 m altitude, which is in turn driven by horizontal pressure gradients and the Coriolis force. The boundary layer consists of three different regions: a very turbulent mixed layer, which transitions into a less turbulent residual layer, and a growing nocturnal boundary layer, which is only sporadically turbulent~\cite{Stull1988}. The mixed layer can be further subdivided into cloud and subcloud layers. While wind turbines operate predominantly in the surface layer up to 100 -- 200 m, which is also denoted as Prandtl layer, AWE systems operate predominantly in the Ekman layer, in which the flow adjusts from the surface layer to the geostrophic wind.

A first global assessment of wind power at high altitudes has been performed by \citet{Archer2009}. The study, based on 28 years of NCEP/DOE reanalysis data, resulted in a global high-altitude wind atlas~\cite{Archer2008} and was one of the scientific drivers for the exploration of airborne wind energy. As part of the study, the optimal harvesting height has been determined, the effect of intermittency has been investigated as well as global climate effects of large-scale energy extraction from higher atmospheric layers. \citet{Miller2011} estimate the maximum sustainable extraction from jet streams of the global atmosphere to be about 7.5 TW and according to \citet{Miller2011b} jet stream wind power does not have the potential to become a significant source of renewable energy.
Using a climate model, \citet{Marvel2013} found that ground-based wind turbines could extract at least 400 TW, whereas high-altitude wind power could extract more than 1800 TW. They further state that uniformly distributed wind turbines generating the entire global primary power demand of 18 TW are unlikely to substantially affect the climate. \citet{Archer2014} explore the global wind power potential by accounting for the specific operational characteristics of AWE systems. Although they allow a variable harvesting height, they also conclude that the optimal locations are where temporally consistent and high wind speeds occur at lowest possible altitudes, to minimize the drag losses of the tether. 

\citet{Gambier2014} and \citet{Gambier2017} present a detailed modeling framework for AWE system designs and combine this with COSMO-EU and NCEP/DOE model data for 12 locations in and around Germany~\cite{Gambier2014} as well as LIDAR measurements up to 1200 m at two locations in Germany~\cite{Gambier2017}. The measurements reveal strong wind shear between 200 and 1000\,m altitude during night time, while during the day the wind shear is small. \citet{Lunney2017} present a techno-economic study of airborne wind energy harvesting in Ireland. The high-altitude wind resource was modeled on the basis of NCEP/DOE AMIP-II Reanalysis (R-2) data which provides an updated 6-hourly global analysis of atmospheric variables such as wind and temperature with $143\times73$ grid points in the horizontal with spacing of 2.5$^\circ$ ranging from the year 1979 to the present. \citet{Yip2017} use MERRA-2 data to identify possible deployment areas for AWE in the Middle East, computing also the optimal height at which the systems would operate. \citet{Emeis2018} discusses AWE systems in his outlook chapter. Only very recently the Dutch Offshore Wind Atlas (DOWA) has been published, consisting of 10 years of hourly data and covering the North Sea Region on a 2.5 $\times$ 2.5 km grid in 17 altitude levels up to 600 m. The data has been obtained by downscaling the ERA5 reanalysis using a regional numerical weather model together with additional satellite and aircraft measurements \cite{DOWA2018}.

\citet{Olauson2018} accurately calculates the wind power generation of several countries and regions using the newly available ERA5 reanalysis data~\cite{ECMWF2018b}. The quality of these predictions and the global availability of an unprecedented spatial and temporal resolution has motivated us to use this data to compare the wind resources available to conventional wind turbines and to AWE systems. While using a fixed harvesting height for wind turbines, we allow AWE systems to continuously adjust the height to the varying wind resource in order to maximize the potential energy yield. The periodic flight trajectory, specific to the different conversion concepts illustrated in Fig.~\ref{fig:concepts}, is not resolved.
Near the coast and onshore, the cost of wind energy is already highly competitive compared to other energy sources. The power generated by conventional wind turbines can vary strongly depending on the consistency of the wind resource. A technology which decreases the variability and increases the available wind power is therefore one of the key measures to allow wind power covering a large part of the base load~\cite{ Archer2009,Zillmann2018}. As part of the objective, the variability of the wind resource is thoroughly investigated. %

The paper is organized as follows. In Section~\ref{sec:data} we briefly introduce the wind data set on which the analysis is based. In Section~\ref{sec:method} we explain the wind resource analysis method and present detailed sample results for a specific location in the English Channel. In Section~\ref{sec:results} the results are discussed for various locations within Europe, followed by the conclusions in Section~\ref{sec:conclusions}.

\section{Data}
\label{sec:data}

In our analysis we cover the area of Western and Central Europe, between
N30.0$^{\circ}$ and N65.0$^{\circ}$ and W20.0$^{\circ}$ and
E20.0$^{\circ}$. The wind resource is estimated using the ERA5 data
provided by the European Centre for Medium-Range Weather Forecasts
(ECMWF) for the period from 1 January 2011 to 31 December
2017. The data is based on a reanalysis of a large set of ground-, air- 
and satellite-based measurements~\cite{ECMWF2018}. The original hourly wind 
data has a resolution of $0.28^{\circ}\times 0.28^{\circ}$ in latitude 
and longitude (ca.~31 $\times$ 31 km). For our current analysis, this data is interpolated 
to a $1^{\circ}\times 1^{\circ}$ (ca.~110 $\times$ 110 km) surface grid. The coarser grid allows displaying regional variations of the wind resource in Europe, but does not resolve local variations.
In vertical direction the data is resolved by 137 barometric altitude levels 
from sea level up to 0.01\,hPa, which is roughly at 80\,km above sea level. 

The sources of the ERA5 data set that are directly relevant for the 
reanalysis of the time-resolved wind speed field are the satellite-based 
measurement data listed in Table~\ref{tab:SatelliteMeasurements} and the 
ground-based measurement data listed in Table~\ref{tab:GroundBasedMeasurements}. 

The documentation of the ERA5 reanalysis data does not provide any information 
about the accuracy of the wind speed estimate. From earlier analyses 
of the ERA-Interim data set and from wind profiler data, the local error of the time-resolved wind speed vector field is estimated to be in the order 
of magnitude of 1\,m/s~\cite{Wentz1992,Lei2016,Andersson2002,ECMWFPrecision,Zhang2018} up to the 850\,hPa level, which is roughly at 1500\,m above sea level. Because we are mainly analyzing harvesting heights up to 500\,m above ground, this altitude range covers most of the analyzed area except for those regions higher than 1000\,m above sea level. 
The precision of the reanalysis data is estimated by comparing it directly with measurements that have and have not been used as input to the reanalysis. The measurements are coherent and, for both cases, no bias of the wind speed is found. 
Most conclusions of our analysis are based on annual averages. The
relatively small estimation errors are both positive and negative and cancel 
each other out as a result of the averaging.
\begin{table}[h]
	\caption{Relevant satellite measurements of the wind speed profile serving as input to the ERA5 reanalysis~\cite{ECMWF2018}.}
	\label{tab:SatelliteMeasurements}
	\begin{center}
		{\scriptsize
			\begin{tabular}{llll}
				\hline
				Microwave sensor & Satellite & Satellite Agency & Data
				Provider \\
				\hline
				MVIRI	       & METEOSAT-2*/3*/4*/5*/7*	       & EUMETSAT/ESA & EUMETSAT \\
				SEVIRI	       & METEOSAT-8*/9*/10	               & EUMETSAT/ESA & EUMETSAT \\
				GOES IMAGER    & GOES-4-6/8*-13*/15*    & NOAA
				& CIMMS*,NESDIS	\\
				GMS IMAGER     & GMS-1*/2/3*/4*/5*	               & JMA & JMA \\	
				MTSAT IMAGER   & MTSAT-1R*/MTSAT2	               & JMA & JMA \\	
				AHI	       & Himawari-8	                       & JMA & JMA \\
				AVHRR	       & NOAA-7 /9-12/14-18, METOP-A  & NOAA &
				CIMMS,EUMETSAT
				\\
				MODIS	       & AQUA/TERRA	                       & NASA &
				NESDIS,CIMMS \\	
				\hline
			\end{tabular}
		}
	\end{center}
\end{table}

\begin{table}[h]
	\caption{Relevant ground-based measurements serving as input to the ERA5 reanalysis~\cite{ECMWF2018,WMO}.}
	\label{tab:GroundBasedMeasurements}
	\begin{center}
		{\scriptsize
			\begin{tabular}{lll}
				\hline
				Data set & Observation Type & Measurements\\
				\hline
				SYNOP                   	& Land station	& Surface pressure, temperature, wind, humidity\\
				METAR                   	& Land station	& Surface pressure, temperature, wind, humidity \\
				DRIBU/DRIBU-BATHY/DRIBU-TESAC	& Drifting buoys& 	10\,m-wind, surface pressure             \\
				SHIP                    	& Ship station	& Surface pressure, temperature, wind, humidity\\
				Land/ship PILOT	                & Radiosondes	& Wind profiles                                \\
				American Wind Profiler  	& Radar	        & Wind profiles                                \\
				European Wind Profiler  	& Radar      	& Wind profiles                                \\
				Japanese Wind Profiler  	& Radar	        & Wind profiles                                \\
				TEMP SHIP               	& Radiosondes	& Temperature, wind, humidity profiles         \\
				DROP Sonde              	& Aircraft-sondes & 	Temperature, wind profiles             \\
				Land/Mobile TEMP         	& Radiosondes	& Temperature profiles                         \\
				AIREP                   	& Aircraft data	& Temperature, wind profiles                   \\
				AMDAR                    	& Aircraft data	& Temperature, wind profiles                   \\
				ACARS                     	& Aircraft data	& Temperature, wind profiles, humidity         \\
				WIGOS AMDAR                 	& Aircraft data	& Temperature, wind                            
				profiles                                     \\      
				
				\hline
			\end{tabular}
		}
	\end{center}
\end{table}

An example of a satellite-based technique that measures the wind velocity vector at the surface is the passive spaceborne microwave radiometry described in~\citet{Gaiser2004}. The technique employs the scattering of sunlight from aerosoles, which creates a polarization of the reflected light similar to the reflection off a reflective surface~\cite{Harmel2013}. For a precision estimate of the input measurements see~\cite{Wentz1992,Lei2016}. 
An example of a ground-based technique that directly measures the wind speed along a certain direction is wind profiling using LIDAR. The technique is based on microwave laser beams that are constantly rotated to different angles with respect to the measurement direction. The Doppler shift and time delay between the laser pulse and the recorded back-scattered light wave allows determining the wind velocity vector from the observations at different angles.

It is expected that the individual wind profiler data listed in Table~\ref{tab:GroundBasedMeasurements}, which is used to extrapolate the ground-based surface wind data to higher altitudes, has a wind speed error below 1\,m/s~\cite{Andersson2002}. In the reanalysis, the historical measurements from all sources in Table~\ref{tab:SatelliteMeasurements} and \ref{tab:GroundBasedMeasurements} are used together with a weather model to estimate how the state of the atmosphere evolved over time. The maximum wind speed error of the reanalysis data is higher than that of the wind profiler measurements because of modeling inaccuracies.%

\section{Method}
\label{sec:method}

In Section~\ref{sec:method:analysis} we first describe the basic resource analysis using the ERA5 wind data set. In Section~\ref{sec:method:potential} we outline the additional wind energy that is accessible by AWE systems and in Section~\ref{sec:method:example} we describe the analysis of the wind potential at variable height using an exemplary location in the English Channel. 

\subsection{Wind Resource Analysis}\label{sec:method:analysis}

The wind resource data in the ERA5 data is specified for barometric altitude levels. Using the ground profile included in the data, we interpolate this data linearly to 25 levels of constant height above ground. This interpolated wind data covers the height range from 10 and 1500 m and is used for the remainder of the analysis.

The measured and reconstructed property is the wind velocity $v$ from which we derive the wind power density
\begin{equation}
\label{eq:windpowerdensity}
P_\text{w} = \frac{1}{2}\rho v^3.
\end{equation}
The air density $\rho$ is approximated by the barometric altitude formula for constant temperature
\begin{equation}
\label{eq:airdensity}
\rho = \rho_0 \exp\left(-\frac{z}{H_\rho}\right),
\end{equation} 
where $\rho_0$ = 1.225~kg/m$^3$ is the standard atmospheric density at sea level and standard temperature, $z$ is the altitude, and $H_p$ = 8.55~km is the scale altitude for density.  

Since energy conversion is generally limited to a lower and upper bound in wind speed---the cut-in and cut-out wind speeds---the shape of the wind speed probability distribution greatly affects the energy yield. To get a better insight into the shape of the distribution at each location, we introduce the 5th, 32nd, and 50th percentiles of the wind speed probability distribution, as also employed by~\citet{Archer2009}. The 50th percentile is equivalent to the median. As shown in Section~\ref{sec:method:example}, the most probable wind speed lies between the 32nd and 50th percentile wind speeds.

\begin{figure}[h]
	\begin{center}
		\includegraphics[width=\textwidth]{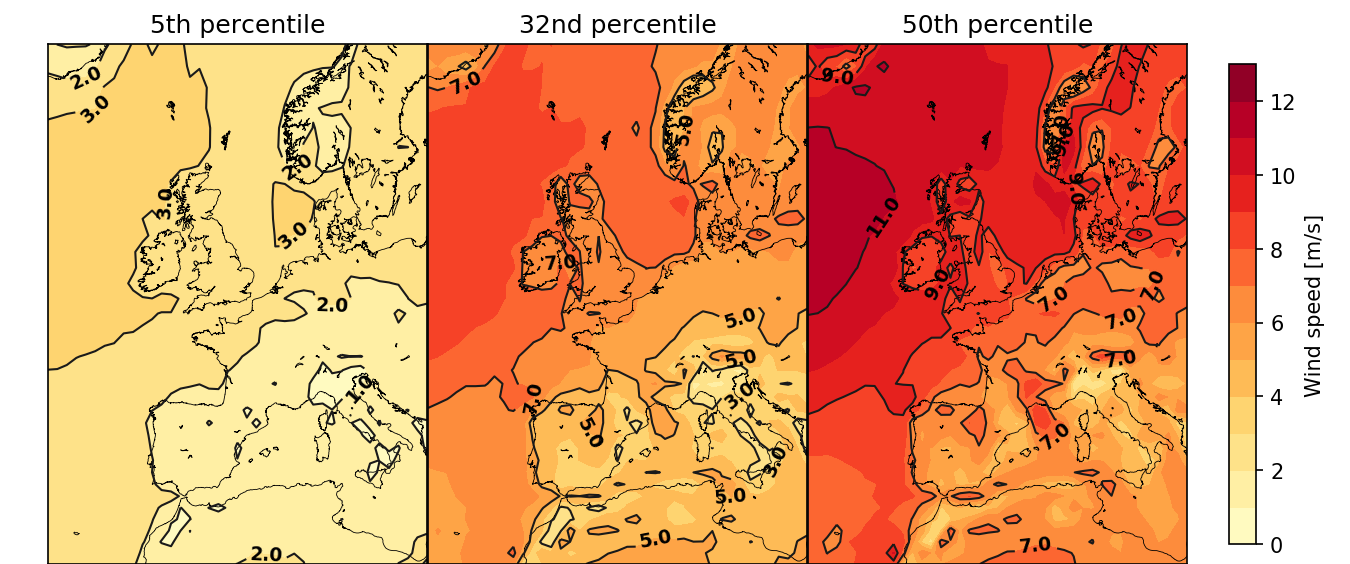}
	\end{center}
	\vspace*{-5mm}
	\caption{5th, 32nd, and 50th percentiles of the wind speed probability distribution at a fixed height of 100 m.}
	\label{fig:fixed_height_wind_plot}
\end{figure}
\begin{figure}[h]
	\begin{center}
		\includegraphics[width=\textwidth]{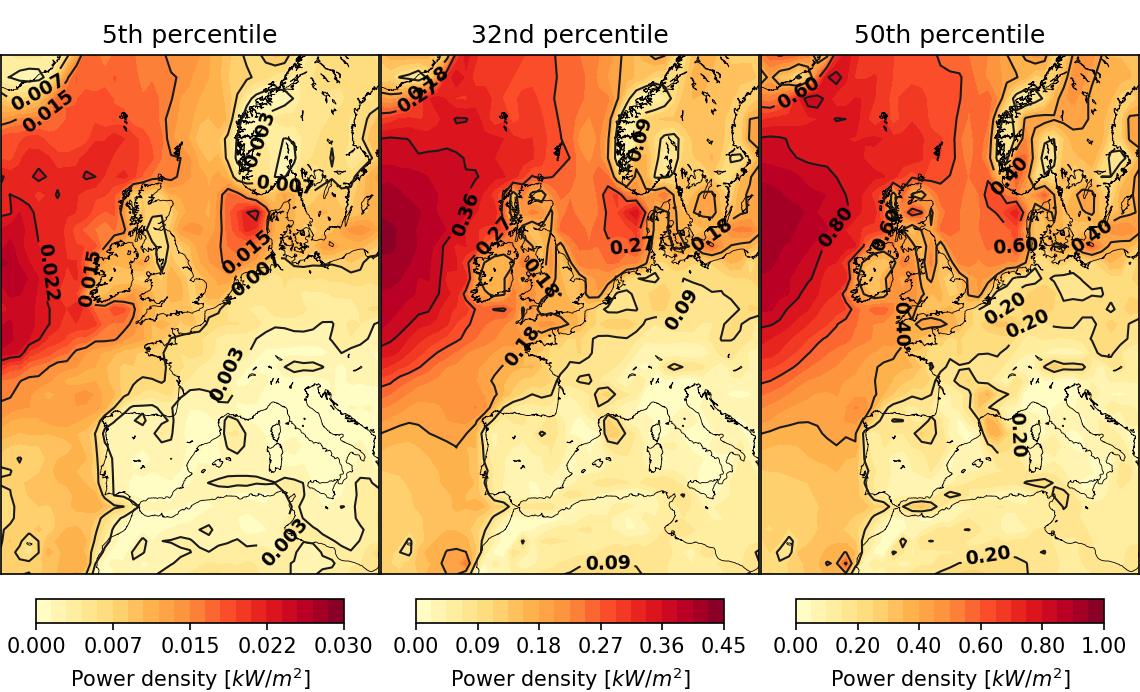}
	\end{center}
	\vspace*{-5mm}
	\caption{5th, 32nd, and 50th percentiles of the wind power density probability distribution at a fixed height of 100 m.}
	\label{fig:fixed_height_power_plot}
\end{figure}

The emphasis in most wind resource analyses is on the height ranges in which conventional wind turbines operate. We illustrate the wind speed and power density probability distributions derived from the ERA5 data at a fixed height of 100\,m in Figures~\ref{fig:fixed_height_wind_plot} and~\ref{fig:fixed_height_power_plot}, respectively. The resulting distributions are used as a reference to assess the wind resource at higher altitudes that are available to AWE systems. We thus denote the 100\,m fixed-height case as \emph{reference case}.


For all three percentiles the highest wind speed is found west of the United Kingdom and the lowest is found south of the Alps. Furthermore, west of the coast of Denmark can be identified as a good offshore site for conventional wind technology. The general trend observed is an increase in wind speed in north-west direction starting from Italy. 

The flight trajectory of a kite for onboard electricity generation, as depicted in Figure~\ref{fig:concepts} (center), is typically circular~\cite{VanderLind2013}. The area swept by the kite can be approximated by an annulus in a plane which is slightly tilted forward with respect to the cross-wind plane. The major difference with respect to the swept area of a conventional wind turbine is that the radius of the annulus can be controlled. The path followed by a kite of a pumping cycle AWE system, shown in Figure~\ref{fig:concepts} (right), is more complex. Because of the superposition of cross-wind motion of the kite and reel-in and reel-out motion of the tether, a larger vertical range is swept, typically between 150 and 500\,m above ground~\cite{Vlugt2013,Vlugt2019,Ruiterkamp2013}. Such a pumping cycle typically takes 1 to 3\,minutes, which is significantly longer than the duration of a single revolution of an onboard generation AWE system.

The height ranges of interest in conventional wind resource analyses are limited and give a poor representation of the resource accessible by AWE systems. As in~\cite{Archer2014} and \cite{Mann2015}, the alternative analysis presented in this paper is not limited to assessing the wind resource at a fixed height, but assesses the potential wind resource when allowing a variable harvesting height. In contrast to \cite{Archer2014}, we however restrict our analysis to height ranges compatible with the technology expected for the first implementations of AWE. In contrast to \cite{Mann2015}, we analyze large geographical areas.%

The analysis is based on two main features of AWE technology: 
\begin{itemize}
	\item the possibility of accessing higher altitudes than tower-based turbines, and
	\item the ability to continuously adjust the harvesting height to the varying wind conditions.
\end{itemize}

The first commercial AWE initiatives envisage a maximum height of 500\,m. Although AWE technology allows access to even higher altitudes, this study focusses on an operational height range between 50 and 500\,m.

The time and length scales of the distinct flow regions in the atmospheric boundary layer depicted in Figure~\ref{fig:atmospheric_boundary_layer} are adequately captured by the temporal and vertical resolution of ERA5. The time scale of the energy conversion process is at maximum in the order of a few minutes and much smaller than the hourly resolution of the wind data. For this reason, the energy conversion process can be regarded as a subscale process and it is thus not resolved in the present analysis. The hourly variation of the optimal height is typically small, as will be shown in Section~\ref{sec:method:example}. Therefore, it is assumed that the AWE system can instantly adjust its harvesting operation within the considered height range.

The length scales of the flight trajectory of a kite and the flow regions in the atmospheric boundary layer are of the same order of magnitude. However, resolving the flight trajectory of a kite is out of the scope of this analysis. The average operational height of the kite is therefore represented by the height which exhibits the maximum wind speed. Based on the earlier description of the flight trajectory of a kite, one could argue that this is a sound assumption for onboard generation AWE systems, but is not necessarily valid for the current pumping cycle AWE systems. Also those could, however, adjust their flight trajectory to a have a larger horizontal and a lower vertical extension, as done in \citet{Ranneberg2018}.

Operating an AWE system at high altitudes allows access to stronger winds, but also comes at a penalty on the energy conversion efficiency. Since the conversion process is out of the scope of this study, we do not account for this efficiency penalty and assume that the height with the maximum wind speed is the optimal operating height for AWE systems.

The software to compile the presented results is implemented in Python and can be downloaded from our publicly accessible repository \citep{Bechtle2018}.
We also provide an archived version of the source code that is packaged together with the original data sets used for this analysis \citep{Schmehl2018b}.
Along with it, we provide instructions on how to download the required ERA5 dataset and how to run the scripts. This will allow future researchers to compile detailed wind statistics at any location in the world. These can be used for analyzing the suitability of specific deployment locations and, in combination with the power curve of a specific AWE system, for computing the annual energy yield.

\subsection{Cumulative High-Altitude Wind Potential}\label{sec:method:potential}

From the wind power density defined in Equation~\eqref{eq:windpowerdensity}, we can derive the mean wind power density integrated over a given height range
%
\begin{equation}
\obar{36}{P^*_{\text{w, $h_0$--$h_1$}}} = \frac{1}{n_\text{hours}} \sum\limits_{i=1}^{n_\text{hours}} \frac{1}{2} \int_{h_0}^{h_1} \rho v^3 dh.
\end{equation}
This property describes the total wind power passing through a vertical strip of unit width, ranging from height $h_0$ to $h_1$ above a given location, and averaged over $n_\text{hours}$ hourly data samples of wind speed and density profiles. It represents the energy potential that is hypothetically available to a conversion technology with the respective harvesting cross section. However, depending of the physical and technical limitations of the specific technology, only a fraction of this hypothetical potential can actually be harvested. For a HAWT the harvesting cross section is the swept area of the rotor, with the well-known physical constraint imposed by the Betz limit~\cite{Burton2011,Kuik2018}. For an AWE system, the harvesting area is variable, as discussed in Section~\ref{sec:method:analysis}, and can generally not be swept as densely as the rotor area of a grond-based turbine, because of the required safety margins to avoid collisions of the flying devices and their tethers~\cite{DeLellis2018}. For this reason, the integral values $\overbar{P_\text{w}^*}$ can not be regarded as an actual metric to compare conventional wind energy and airborne wind energy. We use these values here primarily to highlight the substantial differences of wind power passing through the surface layer and through the entire altitude range of the atmosphere.

\begin{figure}[t]
	\begin{center}
		\includegraphics[width=0.87\textwidth]{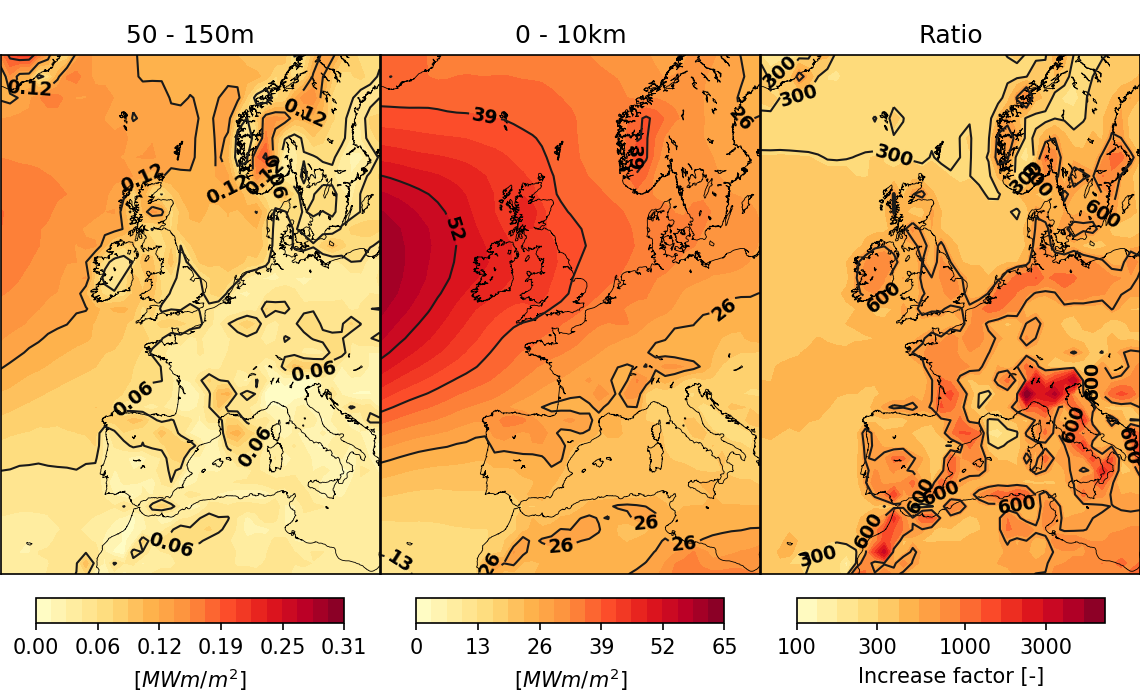}
	\end{center}
	\vspace*{-5mm}
	\caption{Mean wind power density $\overbar{P_\text{w}^*}$ integrated over height for a typical height range 50 to 150 m in which conventional wind turbines operate (left), height range up to 10~km (center), and the respective increase factor $f$ (right). The integral values represent hypothetical potentials that are subject to additional physical and technical limitations of the energy harvesting concepts.}
	\label{fig:integrated_mean_power_plot}
\end{figure}

Figure~\ref{fig:integrated_mean_power_plot} shows the mean wind power density integrated over height for a typical operational height range of conventional wind turbines ($h_0=50$\,m to $h_1=150$\,m) and for the first 10~km above ground ($h_0=0$\,m to $h_1=10$\,km) over Europe, together with the ratio of both values,
\begin{equation}
f = \frac{\obar{51}{P^*_\text{w, 0--10 km}}}{\obar{57}{P^*_\text{w, 50--150 m}}}.
\end{equation}
This ratio quantifies the extend of the untapped wind energy in the lower atmosphere compared to that in only the surface layer.

%

As expected, $\obar{57}{P^*_\text{w, 50--150 m}}$ is significantly higher above the ocean than above land and its isolines follow the coast lines closely~\cite{Sorensen2018}. In contrast to this, $\obar{51}{P^*_\text{w, 0--10 km}}$ does not only show significantly higher power levels, but also a more uniform geographical distribution. The ratio $f$ displayed in Figure~\ref{fig:integrated_mean_power_plot} (right) shows that over the largest part of the sea around Europe, the mean wind power available in the first $10$\,km above ground is 300 times higher than in the layer where wind power is currently harvested by wind turbines. Above the coastal areas, this factor increases to more than $f=600$, in some places exceeding even $f=1000$. This enormous untapped pool of wind energy underscores the importance of investigating the exploitation of higher altitude winds. 

For the reasons stated before, it is clear that only a small fraction of this energy can actually be harvested by AWE systems. Compared to tower-based wind turbines, however, a much larger part of the atmosphere can be accessed while the harvesting height can be adjusted continuously to the varying wind resource and thereby maximizing the energy yield. The following analysis will focus on realistic implementation scenarios of AWE and analyze the realistic gain in wind potential.

\subsection{Detailed Results for the English Channel}\label{sec:method:example}

In this section we detail the analysis for harvesting operation with variable height at a specific fixed location.
Each of the hourly wind profiles is analyzed for the maximum wind speed and the corresponding optimal harvesting height.
The resulting time series are used to determine the wind statistics for the complete investigated period.  

\begin{figure}[h]
	\begin{center}
		\begin{subfigure}[t]{0.49\textwidth}
			\includegraphics[width=\textwidth]{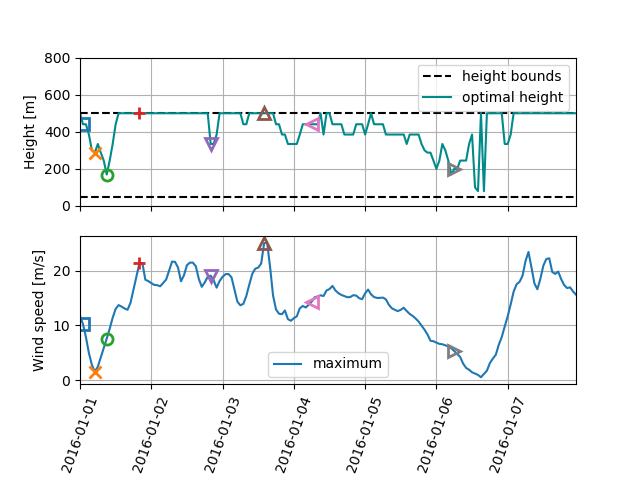}\vspace*{-2mm}
			\caption{Optimal harvesting height over time}
			\label{fig:optimal_height_analysis:optimal_heights}
		\end{subfigure}\hfill
		\begin{subfigure}[t]{0.49\textwidth}
			\includegraphics[width=\textwidth]{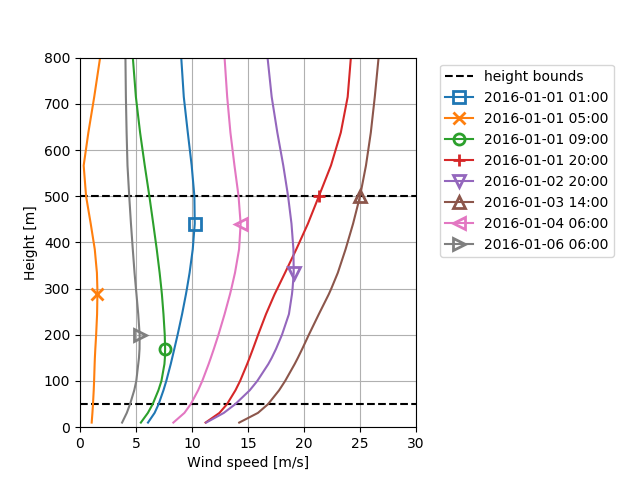}\vspace*{-2mm}
			\caption{Exemplary vertical wind speed profiles}
			\label{fig:optimal_height_analysis:wind_profiles}
		\end{subfigure}
	\end{center}
	\caption{Optimal height analysis at the location N51.0$^{\circ}$,\,E1.0$^{\circ}$ in the English Channel during the first week of 2016.}
	\label{fig:optimal_height_analysis}
\end{figure}

\begin{figure}[h]
	\begin{center}
		\begin{subfigure}[t]{0.49\textwidth}
			\includegraphics[width=\textwidth]{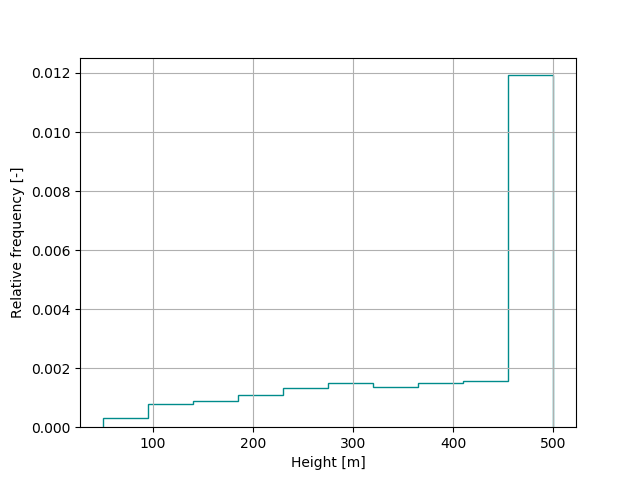}\vspace*{-2mm}
			\caption{Probability distribution of optimal harvesting height}
			\label{fig:windSpeedEnglishChannel:optimal_height_distribution}
		\end{subfigure}\hfill
		\begin{subfigure}[t]{0.49\textwidth}
			\includegraphics[width=\textwidth]{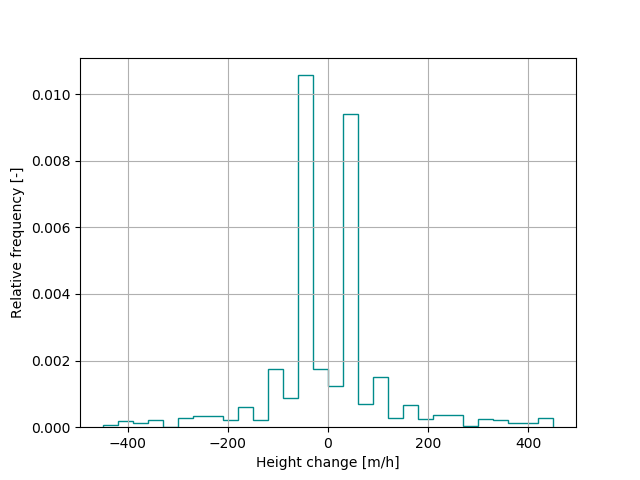}\vspace*{-2mm}
			\caption{Probability distribution of change (hourly) in optimal harvesting height}
			\label{fig:windSpeedEnglishChannel:optimal_height_change_distribution}
		\end{subfigure}
		\begin{subfigure}[t]{0.49\textwidth}
			\includegraphics[width=\textwidth]{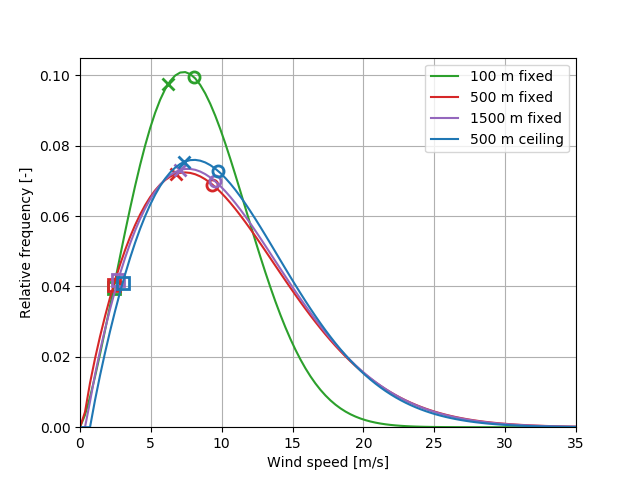}\vspace*{-2mm}
			\caption{Weibull distribution fits for wind speed of fixed-height cases and baseline variable-height cases}
			\label{fig:windSpeedEnglishChannel:wind_speed_histogram2}
		\end{subfigure}\hfill
		\begin{subfigure}[t]{0.49\textwidth}
			\includegraphics[width=\textwidth]{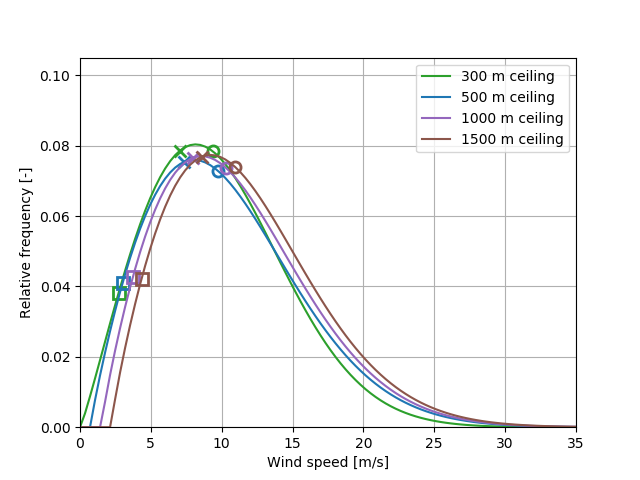}\vspace*{-2mm}
			\caption{Weibull distribution fits for wind speed of variable-height cases}
			\label{fig:windSpeedEnglishChannel:wind_speed_histogram3}
		\end{subfigure}
		\begin{subfigure}[t]{0.49\textwidth}
			\includegraphics[width=\textwidth]{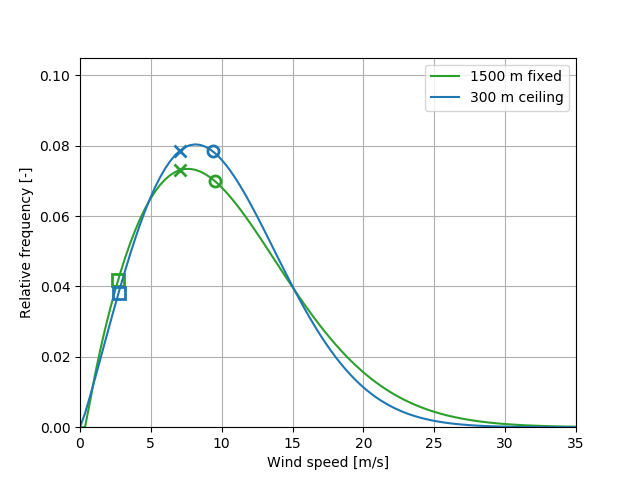}\vspace*{-2mm}
			\caption{Weibull distribution fits for wind speed of 1500\,m fixed-height and 300\,m variable-height cases}
			\label{fig:windSpeedEnglishChannel:wind_speed_histogram3_300m_ceiling}
		\end{subfigure}
	\end{center}
	\vspace*{-4mm}
	\caption{Resulting statistics of the optimal height at the location N51.0$^{\circ}$,\,E1.0$^{\circ}$ in the English Channel for the full data set. In the lower subplots $\Box$, $\times$, and $\ocircle$ refer to the 5th, 32nd, and 50th percentile wind speeds listed in Table~\ref{tab:percentiles_list}.}
	\label{fig:windSpeedEnglishChannel}
\end{figure}

Figures~\ref{fig:optimal_height_analysis} and \ref{fig:windSpeedEnglishChannel} show the wind data at a location in the English Channel. Such an offshore site is in general very suitable for conventional wind turbines~\cite{Sorensen2018}. The optimal harvesting height determined for the first week of 2016 and the corresponding wind speed are shown in Figure~\ref{fig:optimal_height_analysis:optimal_heights}. The markers in the diagram refer to the vertical wind speed profiles shown in Figure~\ref{fig:optimal_height_analysis:wind_profiles}, which exhibit a considerable diversity during this particular week, including a potential low-level jet on 2016-01-02 at 20:00\,h.

\clearpage

\begin{table}[h]
	\caption{5th, 32nd, and 50th percentile wind speeds corresponding to the lines in Figures~\ref{fig:windSpeedEnglishChannel:wind_speed_histogram2} and \ref{fig:windSpeedEnglishChannel:wind_speed_histogram3}}
	\label{tab:percentiles_list}
	\begin{center}
		{\scriptsize
			\begin{tabular}{lp{2.2cm}p{2.2cm}p{2.2cm}p{2.2cm}}
				\hline
				Fixed height & Height [m] & 5th percentile [m/s] & 32nd percentile [m/s] & 50th percentile [m/s] \\
				\hline
				& 100 & 2.4 & 6.2 & 8.0 \\
				& 500 & 2.4 & 6.8 & 9.3 \\
				& 1500 & 2.7 & 7.1 & 9.5 \\
				\hline
				Variable height & Ceiling height [m] & 5th percentile [m/s] & 32nd percentile [m/s] & 50th percentile [m/s] \\
				\hline
				& 300 & 2.7 & 7.1 & 9.4 \\
				& 500 & 3.0 & 7.3 & 9.7 \\
				& 1000 & 3.7 & 8.0 & 10.3 \\
				& 1500 & 4.3 & 8.6 & 10.9 \\
				\hline
			\end{tabular}
		}
	\end{center}
\end{table} 

Figure~\ref{fig:optimal_height_analysis:optimal_heights} shows that the optimal height frequently coincides with the ceiling of the operational range and rarely occurs in its lower region. This observation also holds for the full data set, as can be seen in Figure~\ref{fig:windSpeedEnglishChannel:optimal_height_distribution}. On the considered hourly timescale, the optimal harvesting height is rarely changing and the few variations that do occur are predominately small, as shown in Figure~\ref{fig:windSpeedEnglishChannel:optimal_height_change_distribution}. This gradual variation characteristic justifies the assumption that the harvesting operation can be adjusted instantly.

In a next step, we expand the analysis to the entire 7~year period of the available hourly ERA5 data. For a better comparison, the discrete wind speed distributions are represented by fitted Weibull probability density functions, as illustrated in Figures~\ref{fig:windSpeedEnglishChannel:wind_speed_histogram2}, \ref{fig:windSpeedEnglishChannel:wind_speed_histogram3}~and~\ref{fig:windSpeedEnglishChannel:wind_speed_histogram3_300m_ceiling}. Figure~\ref{fig:windSpeedEnglishChannel:wind_speed_histogram2} shows the differences between the wind speed probability distributions for different fixed-heights cases and the baseline variable-height case with 500\,m ceiling. Within this set of distributions, the 100\,m fixed-height case appears as a distinct outlier. Compared to the distributions at higher altitudes, the distribution at 100\,m is clearly shifted towards lower wind speeds. This is an obvious consequence of the wind shear effect that causes the wind speed to decrease towards the Earth surface, also observable in Figure~\ref{fig:optimal_height_analysis:wind_profiles}. At higher altitudes, the distributions are more similar. Compared to the distribution at 500\,m fixed height, the distribution at 1500\,m is shifted only slightly towards higher speeds. The baseline variable-height case (500\,m ceiling) avoids weaker winds by adjusting the height and virtually never experiences zero wind. The distribution shows a distinct offset from the origin and its centroid shifted farthest towards higher wind speed (higher mean wind speed), both favorable characteristics for a wind speed distribution.

Next to the Weibull functions, the diagrams also include the 5th, 32nd, and 50th percentiles represented by square, cross, and circle markers, respectively. These percentile values are also listed in Table~\ref{tab:percentiles_list}. The percentiles for the actual distributions (shown in the diagrams) agree well with the percentiles for the corresponding Weibull fits. In Table~\ref{tab:comparison_percentiles}, the wind speeds corresponding to the 5th, 32nd, and 50th percentiles of the reference and baseline cases are compared. 
\begin{table}[t]
	\caption{Comparison of the percentiles of the 100\,m fixed-height case and baseline variable-height (500\,m ceiling) case.}
	\label{tab:comparison_percentiles}
	\begin{center}
		{\scriptsize
			\begin{tabular}{lp{2.6cm}p{2.6cm}p{2.6cm}p{2.6cm}}
				\hline
				Percentile & Fixed-height case [m/s] & Variable-height case [m/s] & Absolute increase [m/s] & Relative increase factor [-] \\
				\hline
				5th & 2.4 & 3.0 & 0.6 & 1.27 \\
				32nd & 6.2 & 7.3 & 1.1 & 1.17 \\
				50th & 8.0 & 9.7 & 1.7 & 1.21 \\
				\hline
			\end{tabular}
		}
	\end{center}
\end{table} 
All distributions have their maximum between the 32nd and 50th percentiles (markers $\times$ and $\ocircle$, respectively) which is the range of most probable wind speeds. The ratio of percentiles of the variable-height case and fixed-height percentiles is determined as
\begin{equation}
f_{n^\text{th}-\text{percentile}} = \frac{v_{n^\text{th}-\text{percentile,~variable height}}}{v_{n^\text{th}-\text{percentile,~fixed height}}},
\end{equation}
which we denote in the following as increase factor. Even though this factor is highest for the 5th percentile, the corresponding absolute increase is lowest. The absolute increase shows an increasing trend for an increase in percentile rank. In contrast to this, the increase factor is lowest for the 32nd percentile.

As already mentioned, the probability distribution's centroid is shifted towards substantially higher wind speeds when harvesting above 100\,m. The largest shift is observed when increasing the fixed height from 100 to 500\,m. The shift is less pronounced when switching from the 500\,m fixed-height case to the baseline variable-height (500\,m ceiling) case. Thus, allowing access to winds above 100\,m yields the highest increase in the mean wind speed. This emphasizes the potential of airborne wind energy, even for this site that is very suitable for conventional turbine-based harvesting.

The consistency of the wind is characterized by the 5th percentile of the distribution, where a high value indicates a high consistency. From Table~\ref{tab:percentiles_list} we can see that the 100 and 500\,m fixed-height cases have an identical 5th percentile wind speed. In contrast, we can observe a significant increase from the 500\,m fixed-height case to the baseline variable-height (500\,m ceiling) case. This emphasizes the importance of continuously adjusting the harvesting height to the varying wind conditions for obtaining access to more consistent winds.

Figure~\ref{fig:windSpeedEnglishChannel:wind_speed_histogram3} shows the differences between the distributions for variable-height harvesting with different ceiling heights. Note that the blue curves in Figures~\ref{fig:windSpeedEnglishChannel:wind_speed_histogram2}~and~\ref{fig:windSpeedEnglishChannel:wind_speed_histogram3} are identical, referring to the baseline variable-height case with 500\,m ceiling height. The most pronounced effect of increasing the ceiling height is the increase of the offset between the origin and the distribution, which results in a higher 5th percentile wind speed and thus higher consistency of the wind. Interestingly, the 5th, 32nd, and 50th percentile wind speeds of the 300\,m ceiling variable-height case are virtually identical to those of the 1500\,m fixed-height case. However, the shape and thus the centroid (mean wind speed) of the two distributions are different, as can be seen in Figure~\ref{fig:windSpeedEnglishChannel:wind_speed_histogram3_300m_ceiling}. The two functions intersect at 15\,m/s and the distribution of the 1500\,m fixed-height case has higher probabilities for higher wind speeds.


\section{Results}
\label{sec:results}

In the following we analyze the wind resource for a spatial grid mapping Western and Central Europe. In Section~\ref{sec:results:wind500} we compare the baseline variable-height case with 500\,m ceiling to the 100\,m fixed-height case using contour plots of the 5th, 32nd, and 50th percentiles. In Section~\ref{sec:results:wind1500} we investigate the wind energy potential of future AWE systems with ceiling heights up to 1500\,m.

\subsection{Wind Speed and Power Availability for Variable-Height Operation up to 500\,m}\label{sec:results:wind500}

\begin{figure}[h]
	\begin{center}
		\includegraphics[width=\textwidth]{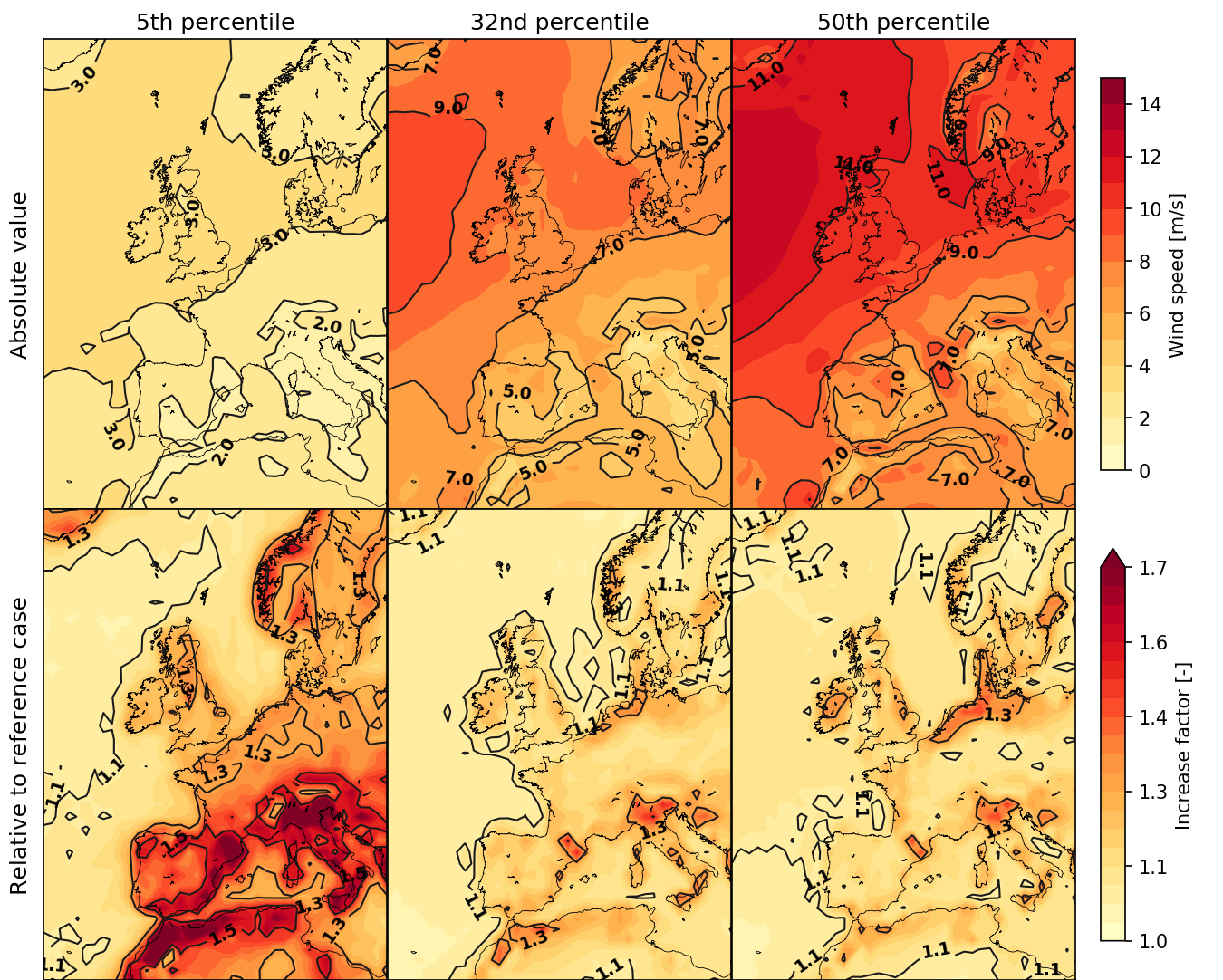}
	\end{center}
	\vspace*{-5mm}
	\caption{5th, 32nd, and 50th percentiles of the wind speed probability distribution for the baseline variable-height (500\,m ceiling) case (top) and the relative increase with respect to the 100\,m fixed-height case (bottom).}
	\label{fig:baseline_comparison_wind_plot}
\end{figure}

The effect of variable-height harvesting on the wind speed distribution is shown in Figure~\ref{fig:baseline_comparison_wind_plot}.
The absolute values (top row) show similar trends as the reference case, e.g. the general trend observed is an increase in wind speed in north-west direction starting from Italy. The bottom row shows the increase factors 
\begin{equation}
\mathrm{f}_{n^\text{th}-\text{percentile}} = \frac{v_{n^\text{th}-\text{percentile,~500\,m~ceiling}}}{v_{n^\text{th}-\text{percentile,~100\,m~fixed}}}.
\end{equation}
The 32nd percentile exhibits the smallest increase in potential wind speed as was also observed for the location in the English Channel. Above the continent, the coastal areas, and the Mediterranean Sea, the increase is more than 10\%, whereas the increase above the Atlantic Ocean appears to be insignificant. The 5th percentile shows the highest increase in potential wind speed. Above coastal areas a 30\% increase is common. Over Italy and the coast of Spain and Northern Africa the increase exceeds 50\%. Peaks in the increase factor of more than 2 are found south of the Alps. Note that these high increase factors coincide with low absolute values in Figure~\ref{fig:fixed_height_wind_plot}. The use of low reference values explains the high relative increase at these sites. As already illustrated in Table~\ref{tab:comparison_percentiles}, the absolute increase at locations with higher reference values is easily underestimated.

\begin{figure}[h]
	\begin{center}
		\begin{subfigure}[t]{0.87\textwidth}
			\includegraphics[width=\textwidth]{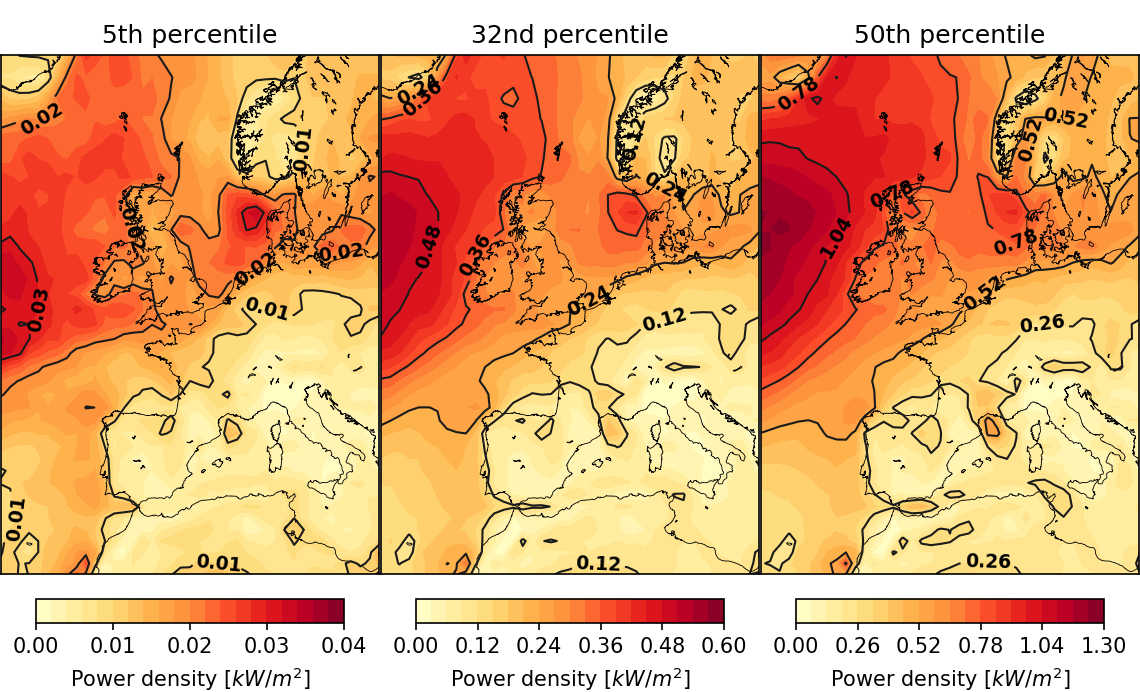}\vspace*{-2mm}
			\label{fig:baseline_power_plot}
		\end{subfigure}
		\begin{subfigure}[t]{0.87\textwidth}
			\includegraphics[width=\textwidth]{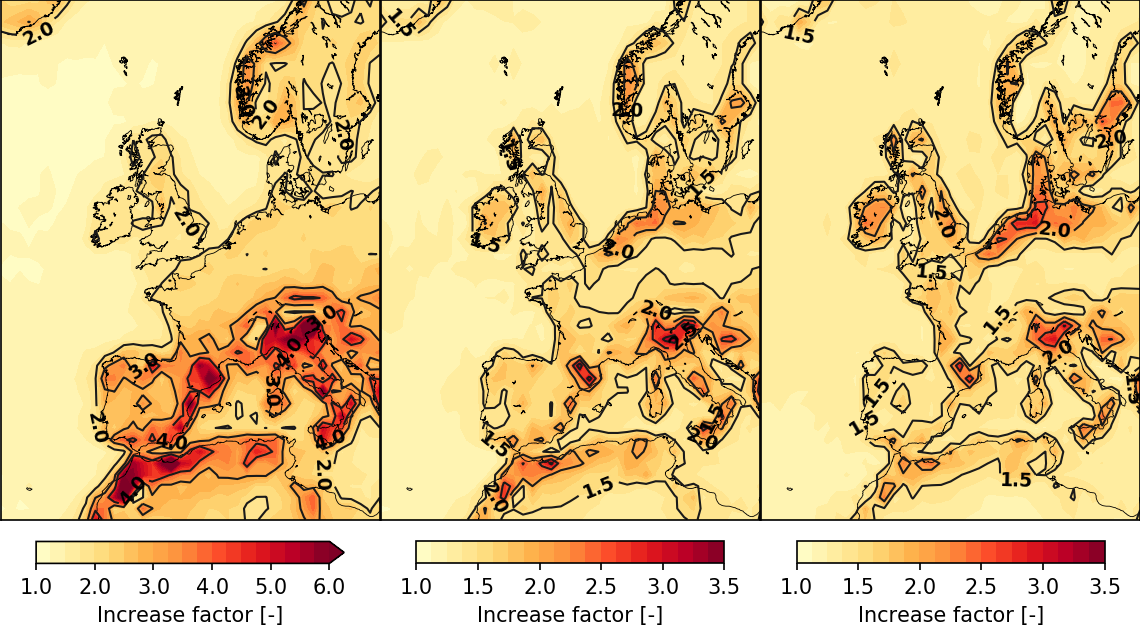}\vspace*{-2mm}
			\label{fig:baseline_power_ratio_plot}
		\end{subfigure}
	\end{center}
	\vspace*{-5mm}
	\caption{5th, 32nd, and 50th percentiles of the wind power density probability distribution for the baseline variable-height (500\,m ceiling) case (top) and the relative increase with respect to the 100\,m fixed-height case (bottom).}
	\label{fig:baseline_comparison_power_plot}
\end{figure}

The wind power density defined by Equation~\eqref{eq:windpowerdensity} is a measure for the wind energy that is locally available for conversion. Figure~\ref{fig:baseline_comparison_power_plot} illustrates the effect of variable-height harvesting on this property. 
The top panels show the geographical distribution of the 5th, 32nd, and 50th percentiles of the wind power density  probability distribution that is available to a typical AWE system operated at variable height up to a ceiling of 500 m. The corresponding percentiles for a fixed height of 100 m, characteristic for conventional wind turbines, are shown in Figure~\ref{fig:fixed_height_power_plot}. The bottom panels of Figure~\ref{fig:baseline_comparison_power_plot} show the geographical distribution of the increase factor, defined as the ratio of wind power densities available to AWE systems and conventional wind turbines
\begin{equation}
f_{n^\text{th}-\text{percentile}} = \frac{P_{\text{w},~n^\text{th}-\text{percentile,~500\,m~ceiling}}}{P_{\text{w},~n^\text{th}-\text{percentile,~100\,m~fixed}}}.
\end{equation}
Note that the color scales differ per plot. Since the wind power density is a function of the cube of the wind speed, the plotted percentile scores and increase factors are roughly the cube of those for the wind speed. Therefore, the same trends hold for the wind speed as for the power density.

To draw rigid conclusions about the wind power availability, the energy conversion process should be taken into account, which is outside of the scope of this study. \citet{Archer2013} argues that the 5th percentile can be used as a generic proxy for the wind power availability. Previously it was shown that allowing a variable harvesting height greatly increases the 5th percentile of both the wind speed and wind power density probability distributions and thereby greatly increase the wind power availability.%

In the following analysis we assess the wind resource by investigating the availability of wind power densities of 40, 300, and 1600\,$\text{W}/\text{m}^2$. The availability $A$ follows from the percentile rank $PR$ corresponding to a wind power density value%
\marginnote{\pdfcomment{
		[MS] For example here it doesn't read well: "wind power availability" and "wind power density availability" are used having different meanings while they only differ 1 word, which can easily be missed by the reader. Furthermore I feel wind power availability is easily mistaken for the equivalent of wind power density: "available wind power".\textCR
		[RS] I replaced "wind power availability" by "wind resource", which is a more generic term and fits well here in my opinion. An alternative would be "resource potential", but I want to limit the use of "potential" because for we acknowledge earlier that for AWE the relationship between the "available" and the actually "extractable" wind power (density) is not so clear yet (the "physical and technical limitations" mentioned earlier). Maybe we should mention the differentiation "available" and the actually "extractable" wind power (density) in the text, for clarity?
}}%
\begin{equation}
\label{eq:availability-percentilerank}
A = 100\% - PR.
\end{equation}
The specified wind power densities roughly correspond to 4, 8, and 14\,m/s wind speed at a height of 500\,m. Note that in the analysis the harvesting height changes over time. Therefore, the wind speeds corresponding to the wind power density values are likely to be slightly different due to the changing air density. The values of 4 and 14\,m/s correspond to typical cut-in and respectively rated wind speeds of conventional wind turbines. Increasing and decreasing the availability of the cut-in and respectively cut-out speed maximizes the operational time of a wind energy system and thereby the wind power availability. The approach of our optimal height analysis is inherent to increasing the availability with respect to the fixed-height reference case. Additionally, the operating height could be adjusted such that not only low wind speeds are avoided, but also high wind speeds. Such an approach allows further tailoring of the wind speed probability distribution to optimize the energy yield of an AWE system, exploiting its ability to adjust the harvesting height to the varying wind conditions, which is out of the scope of this study. The availability of the rated wind speed should be as high as possible to increase the energy yield.

The wind power availability is more precisely predicted using the availability of the cut-in wind speed (or equivalent wind power density) than using the 5th percentile wind speed. Note that the availability of the cut-in wind speed is highly dependent on the magnitude of the cut-in wind speed, which differs substantially from concept to concept. Therefore, the availability of 4\,m/s wind speed might give a false prediction of the wind power availability for some AWE concepts.

\begin{figure}[h]
	\begin{center}
		\begin{subfigure}[t]{0.87\textwidth}
			\includegraphics[width=\textwidth]{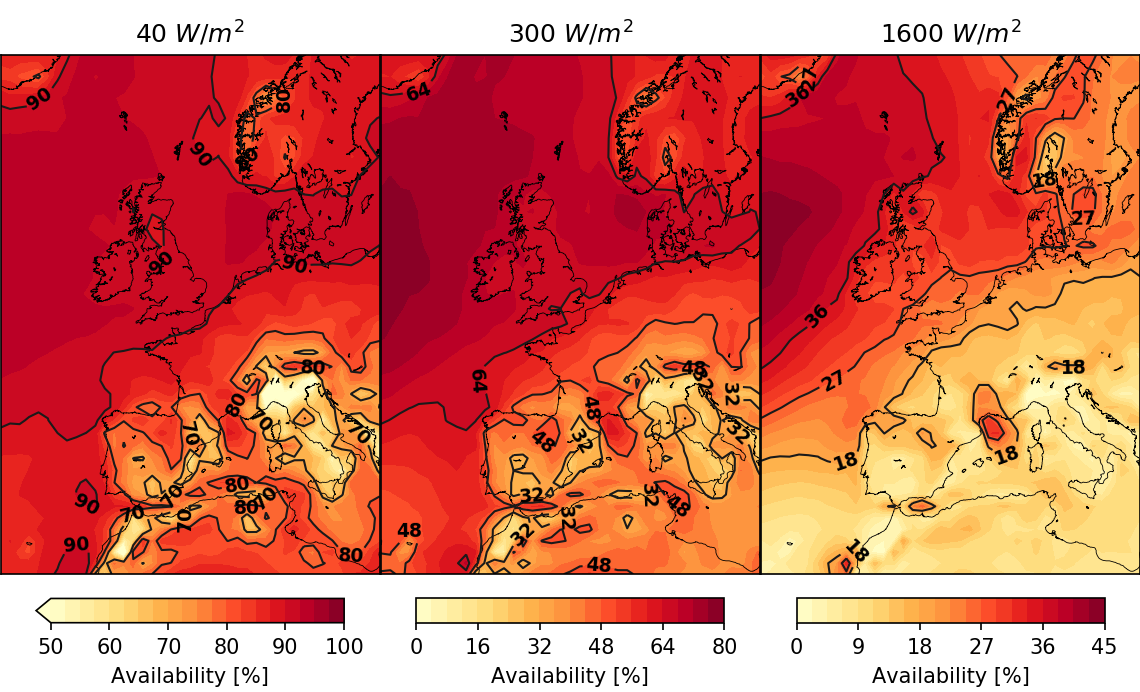}\vspace*{-2mm}
			\label{fig:availability_plot_abs}
		\end{subfigure}
		\begin{subfigure}[t]{0.87\textwidth}
			\includegraphics[width=\textwidth]{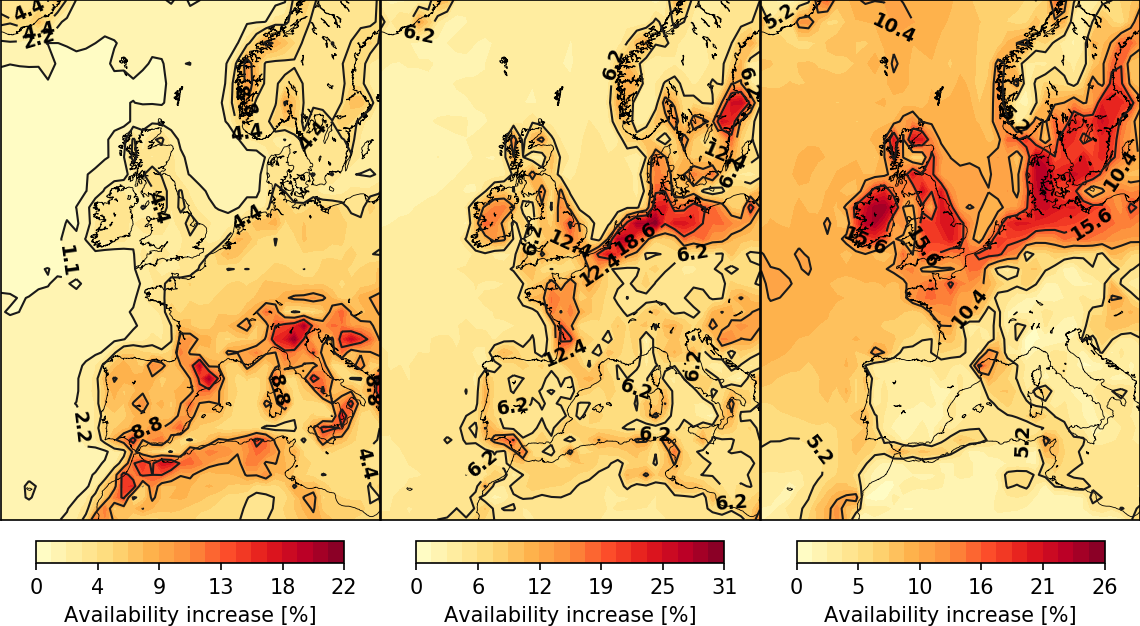}\vspace*{-2mm}
			\label{fig:availability_plot_rel}
		\end{subfigure}
	\end{center}
	\vspace*{-5mm}
	\caption{Availability of 40, 300, and 1600 $\text{W}/\text{m}^2$ wind power density for the baseline variable-height (500\,m ceiling) case (top) and relative increase with respect to the 100\,m fixed-height case (right).}
	\label{fig:availability_plot}
\end{figure}

Figure~\ref{fig:availability_plot} shows the outcome of the wind speed availability analysis. 
Again, the general trend observed is an increase in availability in north-west direction starting from Italy. For 40\,$\text{W}/\text{m}^2$ wind power density, the 90\% availability contour line roughly follows the coast line of Northern Europe. Apparent is the low availability, below 50\%, in the south of the Alps. For the coast line of Northern Europe, the availability of 1600\,$\text{W}/\text{m}^2$ wind power density is roughly 22\%. Also apparent is the high availability at the Mediterranean coast of France. Most of the Mediterranean coastal areas score high on the availability increase
\begin{equation}
\Delta A = A_{\text{500 m ceiling}} - A_{\text{100 m fixed}}
\end{equation}
relative to the reference case for 40\,$\text{W}/\text{m}^2$ wind power density. This is not the case for 1600\,$\text{W}/\text{m}^2$. Because of the high score on the 40\,$\text{W}/\text{m}^2$ availability increase, the overall increase of the wind power availability using AWE is most significant for the Mediterranean coastal areas. The availability increase of 1600\,$\text{W}/\text{m}^2$ wind power density is highest for the United Kingdom, The Netherlands, northern Germany, Denmark, and Sweden. These countries also show a fair score on the availability increase for 40\,$\text{W}/\text{m}^2$. 

Following from the definition in Equation~\eqref{eq:availability-percentilerank}, the contour lines of 50\%, 68\%, and 95\% availability of a certain wind power density in Figure~\ref{fig:availability_plot} correspond to the contour lines of that wind power density in the 50th, 32nd, and 5th percentile panels of Figure~\ref{fig:baseline_comparison_power_plot}, respectively. In general, the availability of a certain wind power density decreases with increasing power density. For example, the availability at the 90\% contour line for 40\,$\text{W}/\text{m}^2$ wind power density in Northern Europe decreases to roughly 60\% and 22\% for 300 and 1600\,$\text{W}/\text{m}^2$, respectively.

The availability increase for 40\,$\text{W}/\text{m}^2$ wind power density emphasizes the potential base load capabilities increase of AWE systems compared to conventional wind turbines. Not only at Mediterranean coastal areas, where the highest availability increase is observed, but over most of Europe the increase is substantial. For instance, over the coast line of Northern Europe the availability is increased by roughly 4.4\%, which is significant considering the already high reference availability of roughly 80\% for conventional wind technology. The increase for 40\,$\text{W}/\text{m}^2$ is mainly realized by the ability of AWE technology to adjust the harvesting height to the varying wind conditions as argued in Section~\ref{sec:method:example}. 


\subsection{Wind Power Availability for Variable-Height Operation with other Ceilings}\label{sec:results:wind1500}

So far, we have only discussed the baseline variable-height case, for which the harvesting ceiling height is set to 500\,m. In a next step, we investigate the effect of the ceiling height on the availability by repeating the availability analysis for 40\,$\text{W}/\text{m}^2$ wind power density. In this section, the availability increase is defined relative to the baseline variable-height case
\begin{equation}
\Delta A = A_{\text{alternative ceiling}} - A_{\text{500 m ceiling}}.
\end{equation}

\begin{figure}[t]
	\begin{center}
		\begin{subfigure}[t]{0.87\textwidth}
			\includegraphics[width=\textwidth]{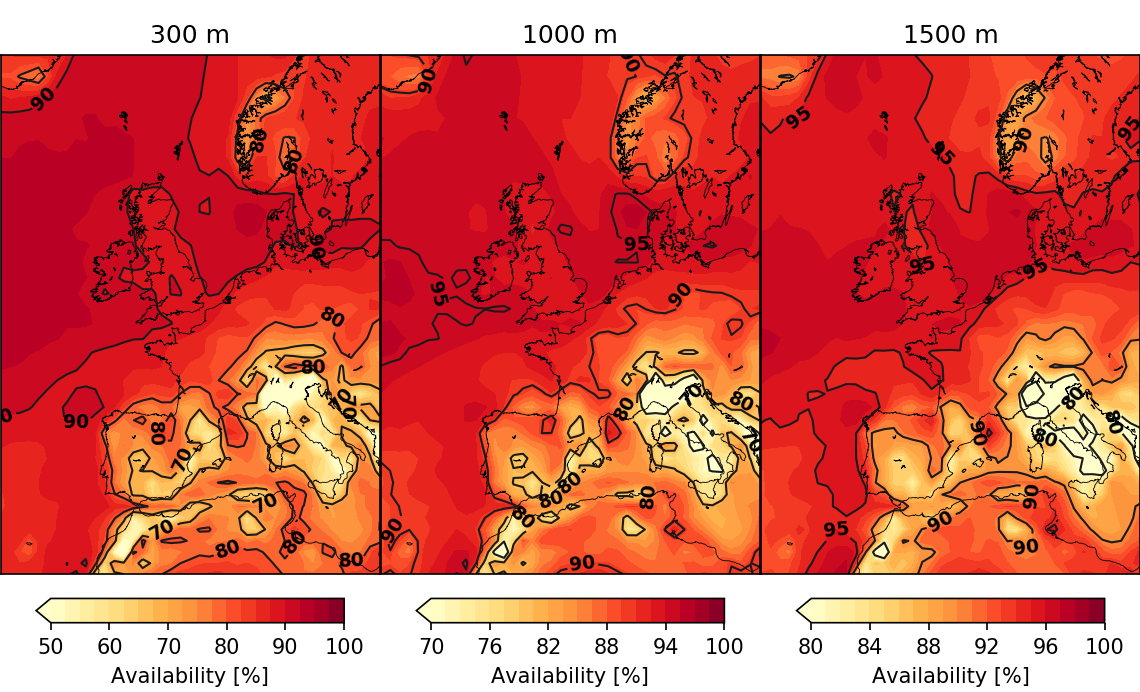}\vspace*{-2mm}
			\label{fig:availability_plot_multiple_ceilings_abs}
		\end{subfigure}
		\begin{subfigure}[t]{0.87\textwidth}
			\includegraphics[width=\textwidth]{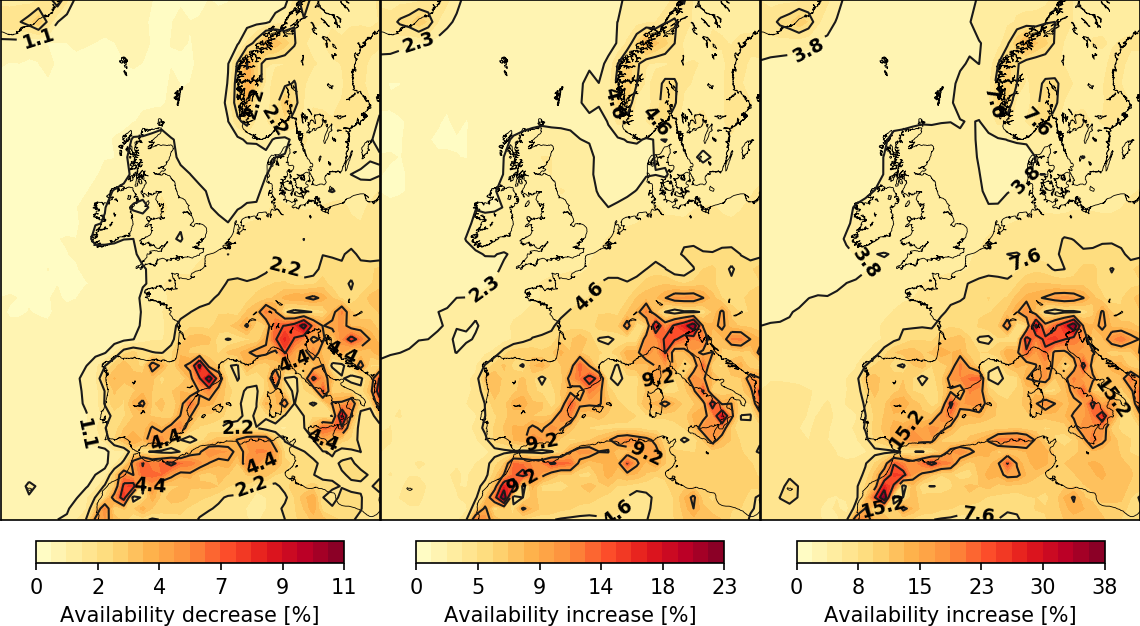}\vspace*{-2mm}
			\label{fig:availability_plot_multiple_ceilings_rel}
		\end{subfigure}
	\end{center}
	\vspace*{-5mm}
	\caption{Availability of 40 $\text{W}/\text{m}^2$ wind power density for variable-height cases with 300, 1000, and 1500\,m ceiling (top) and the corresponding decrease or increase relative to the baseline variable-height case with 500\,m ceiling (bottom).}
	\label{fig:availability_plot_multiple_ceilings}
\end{figure}

Figure~\ref{fig:availability_plot_multiple_ceilings} shows the outcome of this analysis for ceiling heights of 1000, and 1500\,m.
On first sight, the plots for each alternative ceiling height appear to be similar. Note that the color scales differ per plot. For a 300\,m ceiling, the 90\% availability contour line is roughly parallel to the coast line of Northern Europe. A shift of this contour line can be observed when increasing the ceiling height, indicating an expanding area for which 40 $\text{W}/\text{m}^2$ has an availability of at least 90\%. Again, the lowest availability is observed in the south of the Alps, roughly 20\%, 35\%, and 50\% for ceiling heights of 300, 1000, and 1500\,m, respectively (below color scale range). The highest availability can be found west of the United Kingdom, roughly 94\%, 96\%, and 97\% for ceiling heights of 300, 1000, and 1500\,m, respectively.

Figure~\ref{fig:availability_plot_multiple_ceilings} (bottom) shows the decrease or increase with respect to the baseline variable-height case. Note that the availability decreases for the reduced ceiling height of 300\,m and increases for 1000 and 1500\,m ceiling height. The highest increase of wind power availability can be found south of the Alps and in the Mediterranean coastal areas. Note that the areas with the highest availability increase coincide with the areas with the lowest availability, which allow more room for increase. The availability increase of 40\,$\text{W}/\text{m}^2$ wind power density in Figure~\ref{fig:availability_plot} shows the same trend as the availability increase with the ceiling height. In the former, the increase was defined relative to the reference fixed-height case. The effect of switching from a fixed harvesting height to a variable harvesting height on the availability of 40\,$\text{W}/\text{m}^2$ wind power density is thus similar to increasing the ceiling height for variable-height harvesting. The same does not per se hold for higher wind power densities.

As expected, the ceiling height in the variable-height analysis significantly affects the availability of 40\,$\text{W}/\text{m}^2$ wind power density. For example, the contour line crossing the center of France and Germany in the availability increase plots indicate a -2.2\%, 4.6\%, and 7.6\% increase compared to the 500\,m ceiling height case. The decrease is caused by the reduced ceiling height. An increase is mainly realized by the increased minimum probable wind speed for increasing ceiling height, as shown in Section~\ref{sec:method:example}.

\section{Conclusions}
\label{sec:conclusions}

We assess the available wind resources over a large part of Europe using the recently released ERA5 reanalysis data which covers a period of 7 years with hourly estimates at a surface resolution of 31 $\times$ 31 km and a vertical resolution of 137 barometric altitude levels. From this data, we derive 24 levels between 10 and 1500\,m height above ground at a surface resolution of 110 $\times$ 110 km. The analysis is focused on the paradigm of airborne wind energy (AWE): adjusting the harvesting operation up to a predefined ceiling height and allowing access to higher altitudes where winds are generally stronger and more persistent. For the first envisaged AWE systems, a ceiling height of 500\,m is assumed. The operational details, conversion efficiency, and economic boundary conditions vary strongly between different AWE concepts because they are optimized for different conditions and applications. Therefore, in this study, we assess only the potential energy yield and do not account for a specific power conversion mechanism.

The effect of variable-height harvesting is demonstrated for a location in the English Channel. We show the potential of obtaining access to stronger winds by harvesting energy beyond the reach of conventional tower-based wind turbines. We further show that the ability to continuously adjust the harvesting operation to the varying wind conditions is paramount for increasing the minimum probable wind speed and thus obtaining more consistent wind resources.  

The available wind resource for variable-height harvesting with a 500\,m ceiling is analyzed and compared to the wind resource at a fixed height of 100\,m, which represents a typical hub height of a wind turbine. First, the increases of the 5th, 32nd, 50th percentile wind speeds are investigated. The increase of the 5th percentile is most prominent, i.e.{} the wind speed that is exceeded 95\,\% of the time. Over coastal areas a 30\% increase is common. Over Italy and the coast of Spain and Northern Africa the increase exceeds 50\%. The increase is less substantial for the 32nd and 50th percentile wind speeds. 

More relevant for assessing the potential electricity generation is the wind power density and for this reason we repeat the analysis for this flow property. The 5th percentile wind power density increases by more than 100\% over most of Europe compared to the 100\,m fixed-height case. As an alternative we assess the availability of wind power densities of 40, 300, and 1600~$\text{W}/\text{m}^2$. The availability of 40~$\text{W}/\text{m}^2$ approximately describes the percentage of time for which a typical cut-in wind speed of a wind turbine is exceeded. Over most of Europe, the availability is more than 80\,\% for variable-height harvesting with a 500\,m ceiling.

The increase of the 40~$\text{W}/\text{m}^2$ availability emphasizes the increase of base load capability of AWE systems when compared to conventional wind turbines. Not only at Mediterranean coastal areas, where an availability increase of 8 to 10\% is common, but over most of Europe the increase is substantial. Especially considering the high reference availability for conventional wind technology.

Finally, we investigate how ceiling heights of 300, 1000, and 1500\,m above ground affect the availability. As expected, this parameter significantly influences the 40~$\text{W}/\text{m}^2$ availability for variable-height harvesting. The most prominent effect of changing the ceiling height is observed over the Mediterranean coast, but over most of the land mass an increase of 5\,\% or more is observed when increasing the ceiling height from 500 to 1500\,m.%
\marginnote{\pdfcomment{[MS] left out: "This shows that AWE allows access to offshore quality wind almost everywhere in Europe, which would increase the available space for reliable baseload capable wind energy production considerably, and which would shorten supply lines, if the associated technical and legal boundaries of operating an AWE device autonomously close to inhabited areas can be overcome."}}


As a follow-up to the present study, the analysis will be extended to include representative power conversion mechanisms of convention wind turbines and AWE systems. This should shed more light on the Annual Energy Production and base load capability of AWE systems compared to conventional wind turbines. Furthermore, this analysis could be used as a basis in assessing the role of AWE systems in the future global energy mix.

\section*{Acknowledgements}
\label{sec:acknowledgements}
The present analysis is based on data generated using Copernicus Climate Change Service Information
2011-2017. Without the publicly available ERA5 wind data set by the European Centre for Medium-Range Weather Forecasts (ECMWF),
this work would not have been possible. The authors would thus like to express 
their gratitude and support for open data and open science. Mark Schelbergen and Roland Schmehl have received financial
support by the project REACH (H2020-FTIPilot-691173), funded by the
European Union's Horizon 2020 research and innovation programme under
grant agreement No. 691173, and AWESCO (H2020-ITN-642682) funded by
the European Union's Horizon 2020 research and innovation programme
under the Marie Sk{\l}odowska-Curie grant agreement No. 642682.





\bibliographystyle{model3-num-names}
\bibliography{bibliography}







\end{document}